\documentclass[aps,prl,twocolumn]{revtex4-1}
\usepackage{amsfonts,amssymb,amsmath,bm}
\usepackage{graphicx,epsfig,color}

\begin{document}

\title{Inertial floaters in stratified turbulence}

\author{A. Sozza, F. De Lillo and G. Boffetta}

\affiliation{
Department of Physics and INFN, Universit\`a di Torino,
via P. Giuria 1, 10125 Torino, Italy 
}

\begin{abstract}
We investigate numerically the dynamics and statistic of inertial particles 
transported by stratified turbulence, in the case of particle density 
intermediate in the average density profile of the fluid. 
In these conditions, particles tend to form a thin layer around the 
corresponding fluid isopycnal. The thickness of the resulting layer is 
determined by a balance between buoyancy (which attracts the particle 
to the isopycnal) and inertia (which prevents them from following
it exactly). 
By means of extensive numerical simulations, we explore the parameter
space of the system and we find that in a range of parameters particles 
form fractal cluster within the layer.
\end{abstract}

\pacs{47.27.T, 47.55.Hd}

\maketitle

\section{Introduction}
The interaction between particle inertia and turbulence is a fundamental
issue for many problems in natural sciences and in applications, from 
cloud formation in the atmosphere to the dynamics of plankton in the 
ocean \cite{grabowski2013growth,durham2012thin}.
Most of the studies have considered the simplified case of inertial
particles in homogeneous-isotropic turbulence \cite{squires1991preferential,
balkovsky2001intermittent,bec2003fractal}, while more recent works have 
investigated the interaction between gravity and turbulent acceleration
\cite{bec2014gravity,gustavsson2014clustering}.

In the case of stratified turbulence, gravity plays two different roles:
from one side it produces a vertical motion of inertial particles with 
respect to the flow which can increase sedimentation; on the other side
it generates a layered structure in the flow by reducing the
vertical velocity fluctuations and promoting bidimensionalization 
\cite{riley2000fluid,brethouwer2007scaling}.

The motion of inertial particles in stratified turbulence has been 
recently addressed by means of direct numerical simulations within the
Boussinesq approximation of incompressible flow 
\cite{aartrijk2008preferential,aartrijk2010vertical,sozza2016large}.
In particular it has been shown that inertial particles clusterize
for a wide range of parameters in a simplified overdamped limit
in which inertia feels the effect of gravity only.
In that limit, it has been found that vertical confinement due to 
density stratification induces a dissipative dynamics of the particles 
and this produces a strong fractal clustering on the isopycnal surface. 
Fractal clustering depends on a single parameter, combination of the 
Stokes time of the particles and the Brunt-V\"{a}is\"{a}l\"{a} frequency 
of the flow and is maximum for large Stokes numbers \cite{sozza2016large}.

In this Letter we extend the previous investigation 
by considering the Maxey-Riley model for small inertial particles
with finite inertia \cite{maxey1983equation}, similar to the model used in 
\cite{aartrijk2008preferential}.
We confirm the vertical confinement of the particles induced by
stratification and we explain the non monotonic dependence on the 
parameters by introducing a simple stochastic model for the vertical motion.
Small-scale fractal clustering, quantified by the fractal dimension of 
particle distribution, also displays a non-monotonic behavior,
typical of inertial particles in incompressible flows, but with a peculiar
scaling law for small Stokes numbers due to the presence of density 
stratification. 

In this paper we will study the behavior of a widely used model for inertial
particles \cite{maxey1983equation} in order to go beyond the limit of small inertia in
the description of layer formation.  First, we will describe the numerical
model for the flow and the particles and introduce the physical parameters
controlling the dynamics. Then, we will present the numerical results on the
vertical distribution of the floaters and the small scale inhomogeneity.
Finally, the results will be discussed in the conclusions.

\section{Equation of motion for a floater}
We consider a fluid linearly and stably stratified in the 
gravity direction ${\bf g}=(0,0,-g)$ 
with density profile $\rho=\rho_0-\gamma(z-\theta)$. 
Within the Boussinesq approximation the equations of motion 
for the incompressible velocity ${\bf u}$ (with $\bf{ \nabla}\cdot\bf{ u} = 0$)
and the density fluctuation $\theta$ are
\begin{equation}
\partial_t {\bf u} + {\bf u}\cdot{\bf \nabla}{\bf u} = 
-{\bf \nabla}p - N^2\theta\,{\bf \hat{z}} + \nu\nabla^2 {\bf u} + {\bf f},
\label{eq1}
\end{equation}
\begin{equation}
\partial_t \theta + {\bf u}\cdot{\bf \nabla}\theta = 
w + \kappa \nabla^2 \theta ,
\label{eq2}
\end{equation}
where $\nu$ is the kinematic viscosity, $\kappa$ is the diffusivity 
and $N^2=\gamma g/\rho_0$ is the Brunt-V\"{a}is\"{a}l\"{a} frequency. 
In the rhs, ${\bf f}$ represents an external forcing, which injects energy in the system at
a rate $\varepsilon$ to maintain a stationary state, and 
is confined on a characteristic large-scale $\ell_f$.
For simplicity and numerical convenience we only consider the case with Prandtl number 
$Pr=\nu/\kappa=1$.
Three dimensionless parameters can be defined 
in terms of the viscous scale $\eta=(\nu^3/\epsilon)^{1/4}$, 
the Ozmidov scale $\ell_0=(\epsilon/N^3)^{1/2}$ and the forcing scale $\ell_f$:
the Reynolds number $Re=(\ell_f/\eta)^{4/3}$, the Froude Number 
$Fr=\varepsilon^{1/3} \ell_f^{2/3}/N=(\ell_0/\ell_f)^{2/3}$ and the 
buoyancy Reynolds number $Re_b=Fr^2 Re = (\ell_0/\eta)^{4/3}$ \cite{brethouwer2007scaling}.

%
\begin{figure}[t]
\begin{center}
\includegraphics[width=0.49\textwidth]{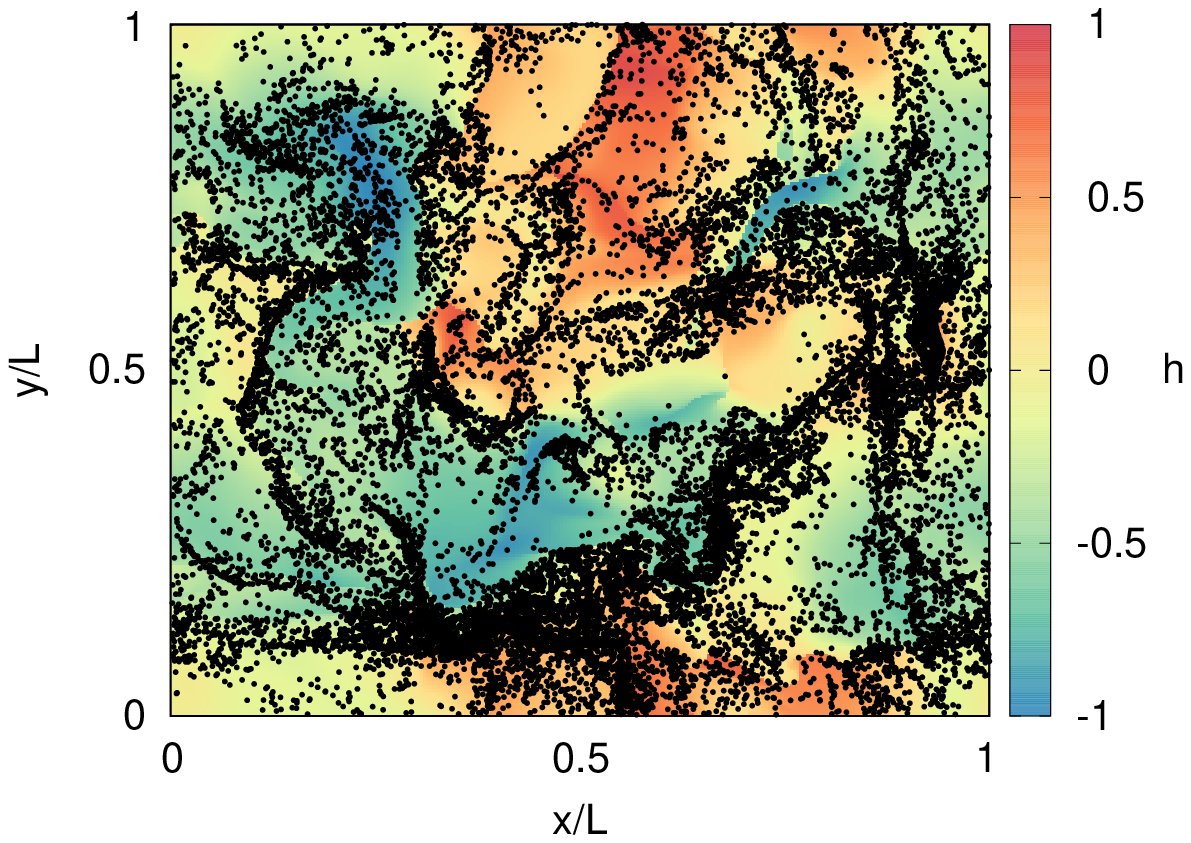}
\includegraphics[width=0.49\textwidth]{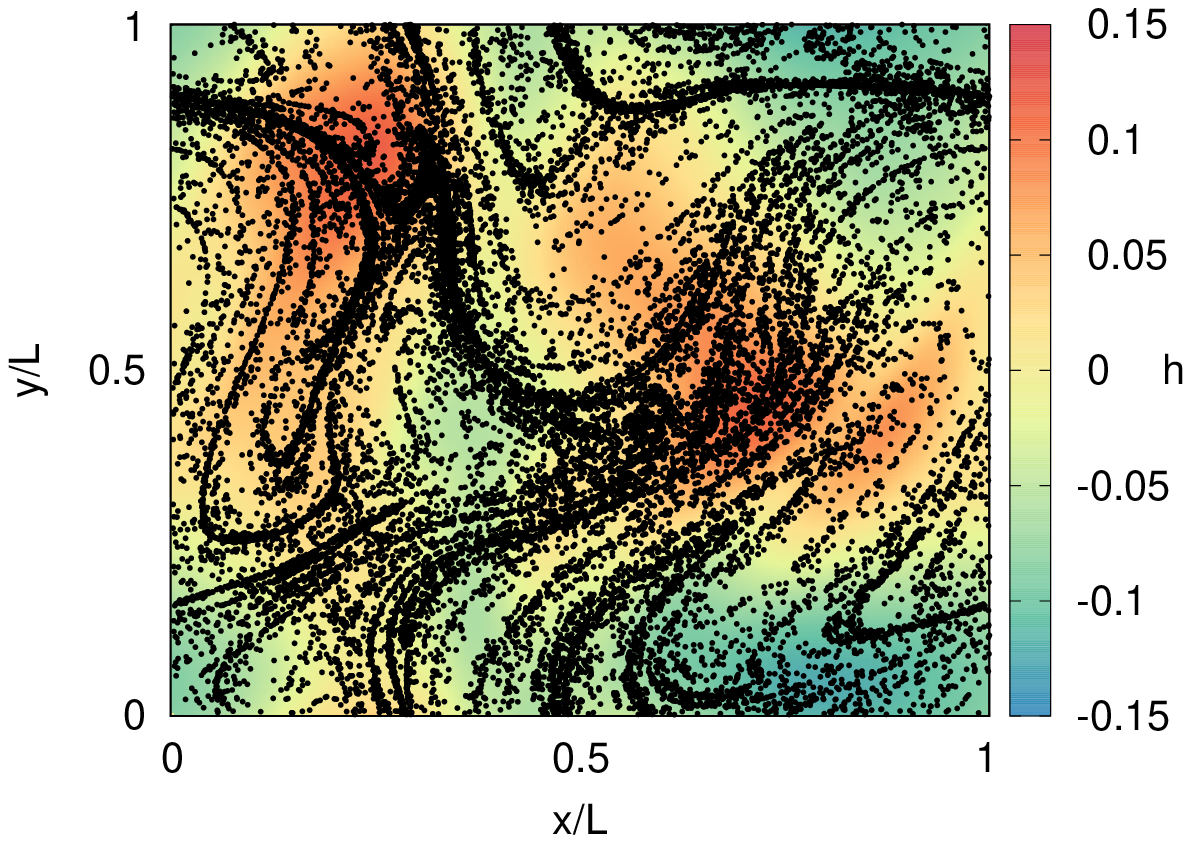}
\caption{(Color online) Snapshots of the distribution of inertial
floaters with $St=2$ over the isopycnal height $h$, implicitly defined by the equation $z-\theta=0$ 
for simulations with $Fr=0.4$ (top panel) and $Fr=0.08$ (bottom panel). Resolution $M=256$.}
\end{center}
\label{fig1}
\end{figure}
%

Particles are assumed to be spherical with a density equal to the
reference density $\rho_0$ of the fluid. This assumption simplifies the notation
without loss in generality, since the flow is homogeneous. We also assume
that their radius $a$ and their velocity relative to the fluid,
${\bf v}-{\bf u}$, are small enough to guarantee creeping flow conditions 
around the particle, i.e. with particle's Reynolds number 
$Re_p=|{\bf v}-{\bf u}| a/\nu \ll 1$. Consistently
with these assumptions, particle position ${\bf x}$ and velocity
${\bf v}$ are governed by a simplified version of the Maxey-Riley equation 
for spherical inertial particles \cite{maxey1983equation}, as
\begin{equation}
\dot{{\bf x}}={\bf v} \, ,
\label{eq3}
\end{equation}
\begin{equation}
\dot{{\bf v}}=\beta\frac{D{\bf u}}{Dt}-
\frac{{\bf v}-{\bf u}}{\tau_p}+\left(1-\beta\right){\bf g} \, .
\label{eq4}
\end{equation}
The added mass term in eq.~(\ref{eq4}) is proportional to the time derivative 
along fluid trajectories (denoted by the material derivative $D/Dt$) of the 
fluid velocity ${\bf u}({\bf x},t)$ at the position of the particle. 
The parameter $\beta=3\rho/(\rho+2\rho_0)$ is a function of the ratio 
between the particle's density $\rho_0$ and the local density $\rho$ of the 
surrounding fluid, while $\tau_p=a^2/(3\nu\beta)$ is the Stokes time.
We remark that in eq.~(\ref{eq4}) we neglect the so-called
Faxen's correction and the Basset's history term since they give
a negligible contribution for small particles.
Taking into account the expression for the density profile, we can
write $(1-\beta)g \simeq {2 \over 3} N^2 (z-\theta)$.
Consistently with the Boussinesq approximation, we further neglect the 
term $(1-\beta)$ when not multiplied by gravity acceleration
to obtain the equation for particle velocity
\begin{equation}
\dot{{\bf v}}={D{\bf u} \over Dt}-
\frac{{\bf v}-{\bf u}}{\tau_p}-{2 \over 3}N^2(z-\theta)\hat{\bf{z}}  \, .
\label{eq5}
\end{equation}
The dimensionless parameter which measures particle inertia is the Stokes
number $St=\tau_p/\tau_\eta$ where $\tau_{\eta}$ is the Kolmogorov
time-scale of the flow.

%
\begin{figure}[t]
\begin{center}
\includegraphics[width=0.49\textwidth]{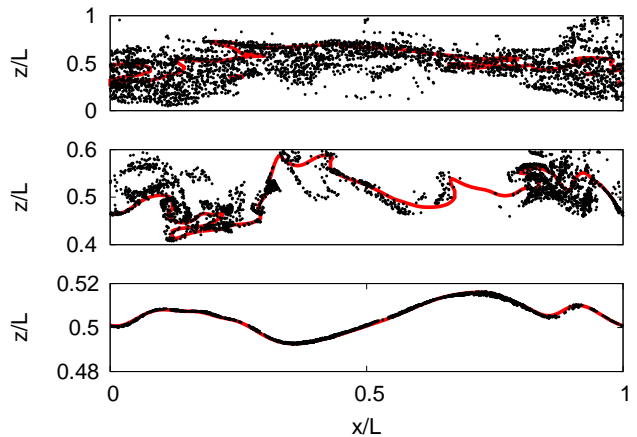}
\caption{(Color online) Snapshots of the vertical distribution of inertial 
floaters with $St=2$ for increasing stratification strength (from top
to bottom $Fr=0.8,0.4,0.08$). Resolution $M=256$.}
\label{fig2}
\end{center}
\end{figure}
%

\section{Numerical Simulation}
We have performed direct numerical simulations of the Boussinesq 
equation for a stratified flow eqs.(\ref{eq1}-\ref{eq2}) in a
periodic box of size $L=2 \pi$
by using a fully-dealiased pseudo-spectral method in space with a 
second-order Runge-Kutta scheme for time evolution. 
Turbulence is maintained in a statistical steady state by a random 
$\delta$-correlated in time forcing acting at large-scale in a narrow band 
of wavenumbers peaked at $k_f \simeq 1/\ell_f$.
We have used different spatial resolutions $M=128,256$, 
corresponding to Reynolds numbers $Re = (k_f \eta)^{-4/3} \sim 390,930$ 
respectively. 
We have considered $4$ different values of stratification with 
 Brunt-V\"ais\"al\"a frequencies corresponding to Froude 
numbers $Fr=0.8,0.6,0.16,0.08$ 
and buoyancy Reynolds numbers $Re_b$ in the range $0.2-35$.
We remark that for this flow the energy is composed by a kinetic and a
potential contributions and therefore the viscous energy dissipation rate 
is typically smaller than the energy input rate $\varepsilon$ 
since a fraction of the kinetic 
energy is converted into potential energy 
and removed from the system through the cascade of potential energy 
\cite{sozza2015dimensional}. As a consequence, small-scale parameters $\eta$ and 
$\tau_\eta$ have a small dependence upon stratification.

Once the turbulent flow has reached a statistical steady state, 
we introduce $20$ populations of $M_p=1.6 \times 10^5$ inertial tracers each,
characterized by different $St$ numbers between $St=0.01$ and $St=100$.
The initial distribution for
each population is uniform in the domain and their trajectories are 
evolved according to eqs.(\ref{eq3}-\ref{eq5}).

Figure~\ref{fig1} plots two snapshots of the particle distribution 
(for $St=2.0$) together with the isopycnal surface $h$, implicitly defined by the equation $z=\theta(\bf{x},t)$
for two values of stratification $Fr=0.4$ (top panel) and $Fr=0.08$ (bottom panel). 
We observe that particles distribution is correlated with the interfacial 
regions between positive and negative values of $\theta$. Such a 
correlation has been observed previously for inertial particles in
homogeneous isotropic turbulence \cite{bec2014clustering}. 
Indeed, once floaters have reached their isopycnal surface, 
they are constrained to perform a quasi-horizontal motion 
which is unaffected by gravity and resembles that of the inertial particles accumulating at the fronts.

The vertical distribution of particles is shown in fig.~\ref{fig2} 
for different values of stratification of the velocity field.
The vertical confinement is produced by the last term in eq.~(\ref{eq5}) which is proportional to $N^2$, 
and it is therefore strongly dependent on stratification. 
For the most stratified case, $Fr=0.08$, 
particles are practically confined on the isopycnal surface $z=\theta$, 
which correspond to the density $\rho=\rho_0$.

%
\begin{figure}[t]
\begin{center}
\includegraphics[width=0.49\textwidth]{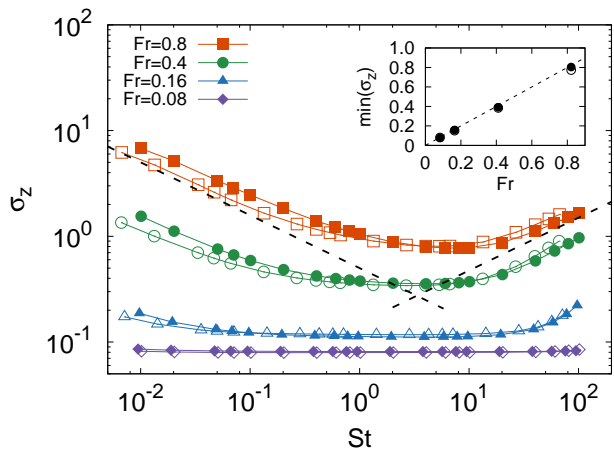}
\caption{(Color online) Layer thickness $\sigma_z$ as a function of $St$ for different stratification 
$Fr=0.8$ (red squares), $Fr=0.4$ (green circles), $Fr=0.16$ (blue triangles), $Fr=0.08$ (purple diamonds) 
and resolution $M=128$ (empty symbols) and $M=256$ (filled symbols).
Dashed lines represents the behavior $St^{1/2}$ and $St^{-1/2}$.
Inset: Behavior of $\sigma_z$ in $Fr$ in the case of maximum confinement.}
\end{center}
\label{fig3}
\end{figure}
%

\subsection{Statistics of vertical confinement}
To better understand the mechanism of layer formation, we can 
first consider the limit of small $St$. For $\tau_p \to 0$,
we can take the overdamped limit of eq.~(\ref{eq5}) 
which, after defining $\tau=3/(2 N^2 \tau_p)$ becomes
\begin{equation}
\dot{\bf x} = {\bf u} - {1 \over \tau} (z-\theta) \hat{\bf z} \, .
\label{eq6}
\end{equation}
The comparison with eq.~(\ref{eq2}) shows that in this limit, 
neglecting diffusion, the quantity $(z-\theta)$ exponentially relaxes to zero 
with the characteristic time $\tau$. Our numerical simulations
confirm this picture, and the variance of the vertical position
$\sigma_z^2=\langle z^2 \rangle - \langle z \rangle^2$ (the 
brackets indicate Lagrangian average) decreases from the initial value
$\sigma^2_0 = \pi^2/3$ (corresponding to uniform vertical distribution) 
and reaches a stationary value after a time of order $\tau$.
In the limit of small $St$, particles distribution is almost homogeneous 
(with $\sigma_z = \sigma_0$), while for increasing $St$, 
$\sigma_z$ drops down indicating that a layer of thickness smaller 
than the box length is formed.

The stationary root mean square of the vertical position as a function of the 
Stokes number and for different Froude numbers is shown in fig.~\ref{fig3}
which confirms the qualitative observation of fig.~\ref{fig2}:
vertical confinement strongly depends on stratification, 
following the fluctuations of the isopycnal surface $ z-\theta = 0$. 
For each $Fr$, we find a maximum confinement 
for intermediate values of the Stokes numbers, $St=O(1)$. 
This behavior can be understood since in the limit of 
small $St$ the last term in eq.~(\ref{eq6}) is small ($\tau$ is large)
and particle trajectory is close to that of a fluid particle. 
In the opposite limit of large $St$, particles are unable
to follow the rapid fluctuations of the isopycnal surface and this 
produces a thick layer around the isopycnal surface.

The above argument can be made more quantitative by considering the 
stochastic version of eq.~(\ref{eq5}) in which 
the fields ${\bf u}$ and $\theta$ are Gaussian noises with zero mean 
and variances $2 D_u$ and $2 D_{\theta}$.
The vertical motion of the floaters is therefore governed by an 
Ornstein-Uhlenbeck process for the probability $p(z,v)$ to find a
particle at the vertical position $z$ with a vertical velocity $v$.
The associated Fokker-Planck equation is \cite{gardiner1985stochastic}
\begin{equation}
\partial_t p + \partial_z (v p) -\partial_v \left[\left({v \over \tau_p} -
{2 N^2 \over 3} z \right)p \right] = D \partial_v^2 p
\label{eq7}
\end{equation}
which has as solution a Gaussian distribution.
The marginal distribution of particle positions is also Gaussian 
with variance $\sigma^{2}_z={3 \tau_p \over 2 N^2} D$, where
\begin{equation}
D= {1 \over \tau_p^2} D_w + {4 \over 9} N^4 D_{\theta}
\label{eq8}
\end{equation}
is the total diffusivity due to $D_w$, the eddy diffusivity due to the
fluctuating turbulent velocity, and $D_{\theta}$, the diffusivity due to the
fluctuations of the stratification (we neglect the contribution in $D$ due to
the added mass since it does not change the following argument).
The expression eq.~(\ref{eq8}) explains the non-monotonic behavior observed
in fig.~\ref{fig3}. Indeed for small $\tau_p$ the first term in the 
r.h.s. of eq.~(\ref{eq8}) dominates
and we get the behavior $\sigma^{2}_z \simeq 3 D_w/(2 N^2 \tau_p) \propto Re_b St^{-1}$ 
(also observed in \cite{birch2009plankton}),
while in the opposite limit we get 
$\sigma^{2}_z \simeq 2 D_{\theta} N^2 \tau_p/3 \propto Re_b^{-1}St$.

Figure~\ref{fig3} shows that $\sigma_z$ strongly depends on the 
stratification, since for small $Fr$ the isopycnal surface to which particles
are attracted is more confined, as shown in fig.~\ref{fig2}. 
This can be also understood from the stochastic model eqs.~(\ref{eq7}-\ref{eq8}) 
in the limit of small $St$ since there we have $\sigma_z \propto 1/N$
which is indeed close to the numerical results as shown in the 
inset of fig.~\ref{fig3} which confirms previous results obtained within
the overdamped model \cite{sozza2016large}. We observe that in the opposite
limit of large $St$, $\sigma_z$ contains the diffusivity $D_{\theta}$ which
is itself dependent on $N$.

%
\begin{figure}[t]
\begin{center}
\includegraphics[width=0.49\textwidth]{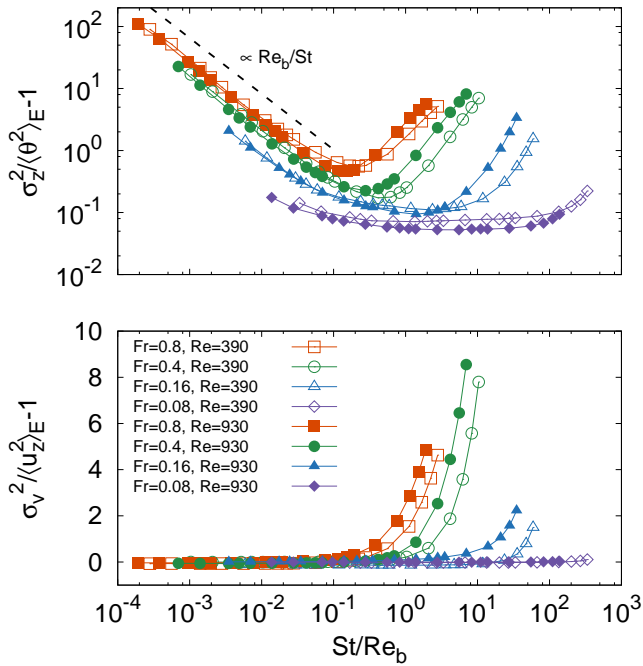}
\caption{(Color online) Variance of the layer thickness $\sigma_z^2$ normalized with the 
Eulerian-averaged variance of the fluctuation density field $\langle\theta^2\rangle$ (top panel) 
and variance of the vertical particle velocity $\sigma_v^2$ normalized with the 
Eulerian-averaged variance of the vertical fluid velocity $\langle u_z^2 \rangle$ (bottom panel)
as a function of $St/Re_b$ for different stratification 
$Fr=0.8$ (red squares), $Fr=0.4$ (green circles), 
$Fr=0.16$ (blue triangles), $Fr=0.08$ (purple diamonds) 
and resolutions $M=128$ (empty symbols) and $M=256$ (filled symbols).
}
\end{center}
\label{fig4}
\end{figure}
%

In fig.\ref{fig4}, the variance $\sigma^2_z$ as a function of $St/Re_b$ 
is shown for different stratifications. For small values $St/Re_b$, numerical results 
confirm the prediction of the stochastic model.

\subsection{Statistics of vertical particle velocity} 
A similar argument using the Fokker-Planck equation eq.~(\ref{eq6}) can be exploited 
to study the vertical particle velocity $v$. 
As before, the marginal distribution of particle velocity is Gaussian 
with variance given by $\sigma_{v}^2 = D_w \tau_p^{-1} + \frac{4}{9}N^4 \tau_p D_\theta$.
In fig.~\ref{fig4} (bottom panel), the vertical particle velocity $v$ is observed to increase with $St$. 
In the limit of small inertia, the particle velocity has to coincide with the velocity of the surrounding fluid.
In the opposite limit, when $St \gg 1$ the vertical particle velocity slightly differs from the fluid velocity and 
it is found to increase approximately as $St^{1/2}$. 
This is very different from the case of inertial particles, 
for which it has been observed that particle velocity decreases as $St^{-1/2}$ \cite{abrahamson1975collision}.

\subsection{Small-scale fractal clustering}
In addition to the vertical, large scale inhomogeneity due to layering around
the equilibrium isopycnal, particle dynamics induces small-scale clustering
within the layer, which is evident from the horizontal section shown in 
fig.~\ref{fig1}.
This small-scale clustering is due to the dissipative dynamics 
eq.~(\ref{eq5}) which governs the motion of the tracers and it is typical 
of inertial particles in incompressible turbulence.
As a consequence, trajectories in phase-space converge onto
a dynamic attractor of smaller dimension and, when the attractor has 
dimension smaller than the space dimension $d=3$, particles distribute
on a fractal set of the same dimension \cite{bec2003fractal}. 

\begin{figure}[t]
\begin{center}
\includegraphics[width=0.49\textwidth]{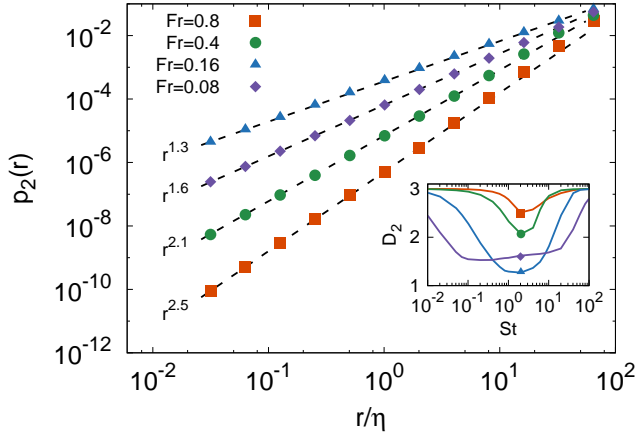}
\caption{Correlation integral $p_2(r)$ as a function of $r/\eta$ for different values of stratification 
$Fr=0.8$ (red squares), $Fr=0.4$ (green circles), $Fr=0.16$ (blue triangles), 
$Fr=0.08$ (purple diamonds)..and Stokes Number $St=2$. Resolution $M=256$. 
Inset: Correlation dimension as a function of $St$ for different stratification.}
\end{center}
\label{fig5}
\end{figure}

In order to characterize small-scale clustering we computed the 
correlation dimension $D_2$ of the particle distribution, defined through 
the probability $p_2(r)\sim r^{D_2}$ of finding two particles within a
distance $r$. A homogeneous distribution in a d-dimensional space would 
clearly give $D_2=d$. 
Figure~\ref{fig5} shows $p_2(r)$ for different values of $St$ and 
$Fr$.  A clear scaling range can be observed between $0.1\eta-10\eta$,
from which one obtains a measure of the correlation dimension $D_2$. The correlation dimension as a function of $St$ and for various values of $Fr$ is plotted in the inset.  In all cases, the strongest
clustering is observed for intermediate values of $St$, while extreme (small or
large) values of $St$ produce more homogeneous layers.
Similar phenomenology is also observed in the small scale
clustering of inertial particles also in homogeneous, isotropic 
turbulence \cite{bec2003fractal}.

\begin{figure}[t]
\begin{center}
\includegraphics[width=0.49\textwidth]{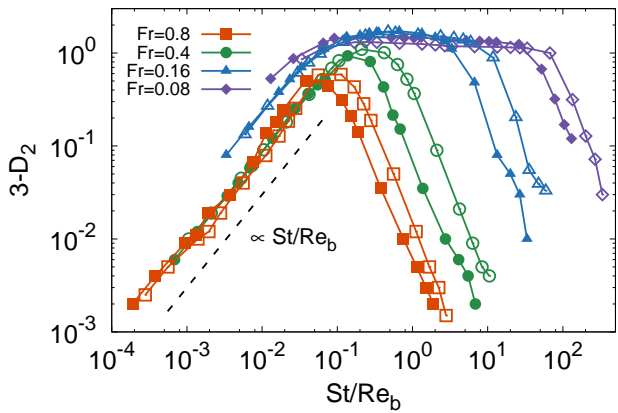}
\caption{Co-dimension $3-D_2$ as a function of $R=\frac{2}{3}Re_b St^{-1}$ 
for different stratification $Fr=0.8$ (red squares), $Fr=0.4$ (green circles), 
$Fr=0.16$ (blue triangles), $Fr=0.08$ (purple diamonds) 
and resolutions $M=128$ (empty symbols) and $M=256$ (filled symbols).
Dashed line represents the prediction $3-D_2 \propto R^{-1}$. 
}
\end{center}
\label{fig6}
\end{figure}

One can focus more precisely on the deviations from homogeneity (for which
$D_2=3$) by considering the co-dimension $3-D_2$, which is plotted in
fig.~\ref{fig6} for the same values of $Fr$. This figure clearly shows a power-law increase of the co-dimension with $St$ in the limit of low inertia.
%
%

%
%

However, comparison between fig.~\ref{fig6} and analogous measurements in the
literature on inertial particles in isotropic turbulence show a striking
difference in the small-$St$ behavior of the curves. 
In the case where buoyancy is neglected one can show that 
$3-D_2\sim St^2$ for $St\to 0$ while in our case 
$3-D_2\sim {St}$ in the limit of small particles, 
as one can see in fig.~\ref{fig6}. 
This behavior can be rationalized following 
\cite{balkovsky2001intermittent,falkovich2002acceleration, fouxon2012distribution}. 
In the small ${St}$ limit, one can again use the approximation of eq.~(\ref{eq6}), in which
the particles are advected by an effective velocity field 
${\bf v}({\bf x},t)={\bf u}-(1/\tau)(z-\theta)\hat{\bf z}$. 
The number density of the particles evolves as 
$dn/dt=-(\nabla\cdot{\bf v})n=(1-\partial_z \theta)/\tau$. 
The particle distribution will be homogeneous above a certain scale 
$\ell \gtrsim \eta$, with a constant number density $n_0$, 
and fractal on smaller scales due to the chaotic dynamics.
If $n$ is sampled with boxes of size $r\ll \ell$, the density fluctuations
measured will be those accumulated along a Lagrangian trajectory during a time
$T_r=\ln(\ell/r)/|\lambda_3|$, i.e. the time it takes for chaotic advection to
compress the initial patch along the direction characterized by
the most negative Lyapunov exponent $\lambda_3$. One therefore has $n^2\sim n_0^2\exp[2\int_0^{T_r}\tau^{-1}(1-\partial_z \theta) dt]$. Eulerian average requires to weigh each patch with its volume, which contracts as $1/n$ so one has 
\begin{equation}
\langle n^2\rangle = n_0^2\left\langle \exp\left[\int_0^{T_r}\tau^{-1}(1-\partial_z \theta) dt\right]\right\rangle
\sim \left(\frac{r}{\eta}\right)^\alpha
\end{equation}
If we assume the integral to be dominated by the constant term we get
$\alpha=1/(|\lambda_3|\tau)$. Of course this requires that
$|\partial_z\theta|\ll 1$, i.e. that the density fluctuations are small. This
condition is typical for $Re_b=O(1)$ or smaller, 
while it is clearly violated around folds in the isopycnal. For 
small $St$ it is reasonable to assume that $\lambda_3\propto
\tau_\eta^{-1}+O(1/\tau)$, so given the definition of the correlation dimension,
$3-D_2=\alpha\simeq \tau_{\eta}/\tau \simeq Re_b^{-1} St$.

It may be interesting to notice that the behavior of $D_2$ is similar to that of $\langle \sigma_z \rangle$ 
in fig.~\ref{fig4}. In particular, both observables have a minimum near $St\sim O(1)$.
Moreover, both quantities shows a plateau in the region of maximum clustering, 
which becomes more evident when the fluid is intensely stratified.\\
Indeed, if $Re_b$ is very small, isopycnals are very close to flat surfaces 
and vertical fluctuations are suppressed.\\

\section{Conclusions}
We investigated the behavior of inertial particles advected by stratified
turbulence, in the case of particle density intermediate within the fluid
density gradient. We have studied how the interplay of gravity and turbulence 
produces a vertical confinement of the inertial particles around the 
isopycnal surface at the particle density, and how the resulting dissipative
dynamics produces fractal clustering at small scales.

The interaction between particle settling and turbulence is an essential
problem for many systems in the natural sciences and in applications. In
marine and lake biology, the possibility of organisms to control their position
in the water column is of great importance for the uptake of nutrients, the
access to light and for escaping predators \cite{reynolds2006,kiorboe2008}.
In the absence of specific mechanisms, the survival of such organisms must
result from the complex interplay between turbulent mixing and growth
\cite{huisman2002how,margalef1978life}.
A particularly interesting problem is posed by thin phytoplankton layers (TPL),
aggregations of phytoplankton and zooplankton at high concentration, with
thickness from centimeters to few meters, extending up to several kilometers
horizontally and with a time scale from hours to days \cite{durham2012thin}.
Various mechanisms have been proposed to explain such formations, notably
depending on the motility (or absence thereof) of a particular species observed
to form TPLs. In particular, If a cell is neutrally buoyant at some
intermediate depth in the pycnocline, then buoyancy could lead to the formation
TPLs \cite{alldredge2002occurrence, durham2012thin,sozza2016large}.

The physical and biological implications of fractal clustering observed
at small scales are associated with an increased probability to find 
particles at small separations with respect to a homogeneous distribution. 
Numerical studies inspired by TPLs showed that this can have important effects on the population dynamics of plankton \cite{perlekar2010}.
In general, small scale clustering is relevant in problems where encounter rates are important, such as mating and competition for resources in biology, coalescence and coagulation of droplets in cloud physics \cite{falkovich2002acceleration,shaw2003} and engineering applications \cite{post2002}.  

We thank M. Cencini for useful discussions. We acknowledge support from the 
COST Action MP1305 "Flowing Matter" and from Cineca within the INFN-Cineca agreement INF17turb.


%

\end{document}